# VO$_2$ microcrystals as advanced smart window material at semiconductor to metal transition

Raktima Basu,[1,*] P. Magudapathy,[2] Manas Sardar,[2] Ramanathaswamy Pandian,[1] Sandip Dhara[1,*]

[1] Nanomaterials Characterization and Sensors Section, Surface and Nanoscience Division, Indira Gandhi Centre for Atomic Research, Homi Bhabha National Institute, Kalpakkam- 603102, India

[2] Materials Physics Division, Indira Gandhi Centre for Atomic Research, Kalpakkam-603102, India

*Abstract* - Textured VO$_2$(011) microcrystals are grown in the monoclinic, M1 phase which undergo a reversible first order semiconductor to metal transition (SMT) accompanied by a structural phase transition to rutile tetragonal, R phase. Around the phase transition, VO$_2$ also experiences noticeable change in its optical and electrical properties. A change in color of the VO$_2$ micro crystals from white to cyan around the transition temperature is observed, which is further understood by absorption of red light using temperature dependent ultraviolet-visible spectroscopic analysis and photoluminescence studies. The absorption of light in the red region is explained by the optical transition between Hubbard states, confirming the electronic correlation as the driving force for SMT in VO$_2$. The thermochromism in VO$_2$ has been studied for smart window applications so far in the IR region, which supports the opening of the band gap in semiconducting phase; whereas there is hardly any report in the management of visible light. The filtering of blue light along with reflection of infrared above the semiconductor to metal transition temperature make VO$_2$ applicable as advanced smart windows for overall heat management of a closure.

Keyword: vanadium dioxide, SMT, SPT, smart window, Raman spectroscopy

[*]Corresponding Authors E-mail: raktimabasu14@gmail.com; dhara@igcar.gov.in

# 1. Introduction

Vanadium dioxide ($VO_2$) is well known for its first order semiconductor to metal transition (SMT) between low-temperature monoclinic to high-temperature rutile tetragonal phase at a technologically important temperature of 340K [1-3]. Along with the phase transition, significant changes also take place in the optical and electrical properties of the material [4,5]. The resistivity changes by four orders of magnitude through the transition [6], and the material reflects the electromagnetic radiation in the infrared (IR) region at high temperature metallic phase, whereas it transmits IR at low temperature semiconducting phase [4,7,8]. However, the material is known to be transparent in the visible region [9]. Generally, the optical property of a material depends on the response of electron and its transition between electronic states due to perturbation by the incident radiation. As the SMT in $VO_2$ is also accompanied by a structural phase transition (SPT), there have been controversies on the driving mechanism of the phase transition in $VO_2$; whether it can be understood in the Peierls scenario, or by invoking electronic correlation, indicating decisive role played by Mott physics. According to Zylbersztejn and Mott [10], Sommers and Doniach [11], and Rice *et al.* [12] Coulomb repulsion is responsible for opening the energy gap in the semiconducting phases of $VO_2$. Qualitative proposal of electronic structures of $VO_2$ were reported long ago by Goodenough [13]. The $O2p$ orbitals stay 2.5 eV below the Fermi level in the electronic band structure of $VO_2$, [14,15] and form the valence band with π and σ bonds. In $VO_2$, there is one *d* electron per V atom and the *d* levels of the V ions split into two states, namely, the lower lying triply degenerate $t_{2g}$ and higher lying doubly degenerate $e_g^\sigma$ states. The $t_{2g}$ multiplet splits again into an $a_{1g}$ state ($d_{xy}$) and an $e_g^\pi$ ($d_{xz}$, $d_{yz}$) doublet due to the tetragonal crystal field [13]. In the low temperature semiconducting phase, V atoms form pairs in the rutile *c* ($c_R$) direction and split the $a_{1g}$ bands into lower (bonding, $a_{1g}$) and upper (antibonding, $a_{1g'}$) bands. Moreover, the V-V pairs twist and thereby enhance the V$d$-O$p$ hybridization, which leads to rise in the $e_g^\pi$ band above the Fermi level. Thus, a gap of ~ 0.7 eV opens up to stabilize the insulating phase [13,16]. It is important to note that the density functional theory (DFT) with local density approximation (LDA) [16], however, fails to open the band gap between $a_{1g}$ and $e_g^\pi$ bands. On the other hand, LDA along with cluster dynamical mean field theory (C-DMFT) does manage to open up an energy gap of correct magnitude by taking account the correlation effects (Hubbard *U*) [17]. However, in the metallic phase, the energy gap collapses and the Fermi level crosses partially filled $a_{1g}$ and $e_g^\pi$ bands. The optical properties around the transition temperature are studied mostly in the IR region, which supports the opening of the band gap in semiconducting phase; whereas there is hardly any report in the visible region.

In the present report, we observed a change in color of the $VO_2$ micro crystals from white to cyan around the transition temperature, which is further confirmed by absorption of red light and reflection of blue light using temperature dependent ultraviolet-visible (UV-Vis) spectroscopic analysis and photoluminescence (PL) studies. The absorption of red light is explained by the optical transition between Hubbard states, supporting the electronic correlation as the driving force for SMT in $VO_2$. The filtering of blue light at semiconductor to metal transition in $VO_2$, which is also well known for IR reflector at high temperature metallic phase, makes $VO_2$ applicable as smart windows for overall heat management of a closure.

## 2. Experimental details

$VO_2$ micro-crystals were synthesized on Si (111) substrate by vapor transport process using Ar as carrier gas. Bulk $VO_2$ powder (Sigma-Aldrich, 99%) was used as source and was placed in a high pure (99.99%) alumina boat inside a quartz tube. The reaction chamber was kept in a furnace and evacuated up to $10^{-3}$ mbar. Si (111) substrate was kept 1 cm away from the source and normal to the Ar flow direction. The synthesis was carried out at 1150K for 3 h. The temperature of the quartz tube was programmed to rise up to the optimized growth temperature with a ramp rate of 15K min$^{-1}$. The morphology of the pristine sample was analyzed using a field emission scanning electron microscope (FESEM, SUPRA 55 Zeiss). The pure $VO_2$ phase and crystallographic analysis were studied by X-ray diffraction measurements with the help of glancing incidence x-ray diffractometer (GIXRD; Inel, Eqinox 2000) with a glancing angle ($\theta$) of $0.5^o$ in the $\theta$-$2\theta$ mode using a Cu K$_\alpha$ radiation source ($\lambda$=1.5406 Å). The vibrational modes of the grown sample were studied by Raman spectroscopic analysis using a micro-Raman spectrometer (inVia, Renishaw, UK) in the backscattering configuration with Ar$^+$ Laser (514.5 nm) as excitation source, diffraction gratings of 1800 gr.mm$^{-1}$ and a thermoelectrically cooled CCD camera as the detector. Electrical measurements were carried out in a voltage range of 2 V using two Au coated contact tips as work function of Au (5.1 eV) matches with that of $VO_2$ (~ 5.15-5.3 eV depending on the metallicity) [18]. The temperature dependant resistance values are recorded by keeping the sample in a temperature controlled stage (Linkam; THMS600) with the help of a source measurement unit (Agilent B2911A). The optical images of the samples were captured at different temperature with help of the same optical microscope using long working distance 50X objective with numerical aperture (N.A.) of 0.45. The images are in RGB format with spectral response (gamma) and light intensity corrected. The PL spectra were collected at different temperatures using the same micro-Raman spectrometer with 514.5 nm Laser excitation. Temperature dependent absorption spectra were recorded using an UV-Vis absorption spectrometer (Avantes) in reflection geometry with a light source of excitation wavelength in the range of 200 to 800 nm. A bare Si wafer was used for reference to nullify the reflected contribution from the substrate. In order to perform the temperature dependent spectroscopic measurements, the samples were kept in a Linkam (THMS600) stage, driven by an auto-controlled thermo-electric heating and cooling function within a temperature range of 300 to 360K.

## 3. Results and discussion

### 3.1. Morphology and structural studies

The typical FESEM image of the grown microcrystals is shown in Figure 1(a). The average diameter of the microcrystals is 5 to 10 μm. The crystallographic structural study (Fig. 1(b)) shows (011) plane corresponding to the monoclinic M1 phase of $VO_2$ (JCPDS # 04-007-1466). The single peak at 2θ = 27.98° in the GIXRD spectra confirms the (011) textured microstructures.

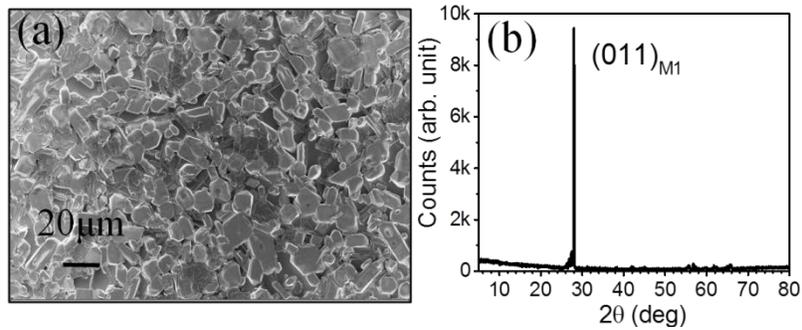

**Figure 1.** (a) FESEM image of as-grown microcrystals and (b) GIXRD spectrum of as-grown sample indicating crystalline (*hkl*) planes of (011) corresponding to M1 phase of $VO_2$

### 3.2. Vibrational analysis and in-situ optical imaging

Among Group theoretically predicted eighteen Raman active modes for M1 phase of $VO_2$ at $\Gamma$ point, $9A_g+9B_g$ [19], we observed twelve modes as shown in figure 2.

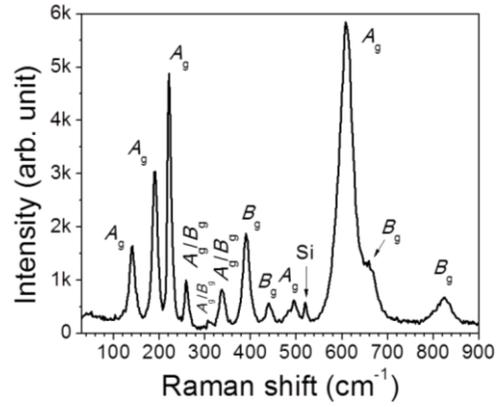

**Figure 2.** Typical Raman spectrum of the as grown $VO_2$ microcrystals with excitation of 514.5 nm at room temperature.

Raman modes at 140($A_g$), 190($A_g$), 221($A_g$), 258 (either $A_g$ or $B_g$,; $A_g/B_g$), 307($A_g/B_g$), 339($A_g/B_g$), 391($B_g$), 427($B_g$), 501($A_g$), 611($A_g$), 670($B_g$), 823($B_g$) cm$^{-1}$ confirm the presence of pure $VO_2$ in the M1 phase [20,21]. In order to study the phase transition in $VO_2$, temperature dependent Raman spectral measurements were performed within a temperature range of 300 to 360K. Figure 3(a) shows the temperature dependent Raman spectra for both increase and decrease in temperature.

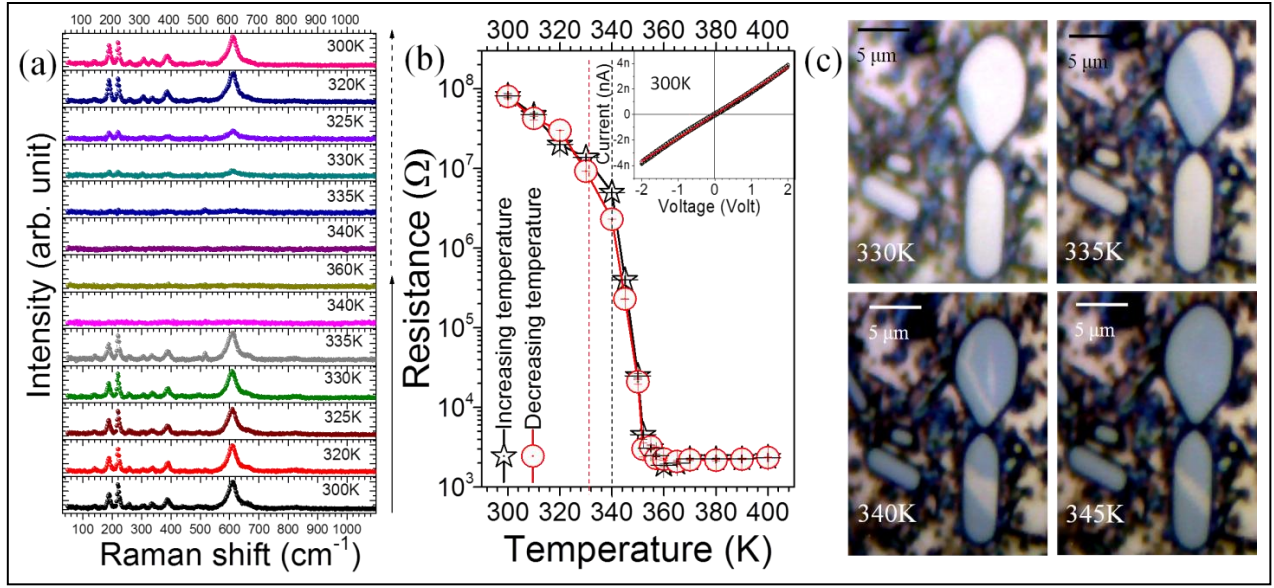

**Figure 3.** (a) Raman spectra of the sample at the temperature range 300 to 360K. Solid and dashed arrows denote increasing and decreasing temperatures, respectively, (b) Reversible temperature dependent resistance measurement for the $VO_2$ crystals showing a change in resistance of four order indicating metal insulator transition with a hysteresis of ~8K (shown by vertical dashed lines). Inset shows linearly fitted (solid line) reversible IV curves at 300K confirming Ohmic behavior (c) optical images of $VO_2$ micro-crystals at different temperatures.

Temperature dependent Raman spectroscopic study shows that all Raman modes disappear at 340K (while increasing the temperature), confirming the SMT and reappear again at 330K (while decreasing the temperature), as shown in figure 3(a). A hysteresis of value ~10K is observed. A first order and reversible phase transition close to room temperature is reported for $VO_2$ with a hysteresis of value ranging from 3 to 40K depending on doping, size distribution, nucleation as well as crystallographic orientation [1-3]. Temperature dependent resistance measurement was carried out to study the phase transition in $VO_2$. A sharp drop in resistance of four orders is observed at 340K confirming the MIT in $VO_2$ (Figure 3(b); inset showing Ohmic contact) with a hysteresis of value ~8K, which matches with the observed hysteresis ~10K in the Raman studies. The optical images of the samples were captured at different temperatures using optical microscope with 50X objective with numerical aperture (N.A.) of 0.45 using white light source. A change in color of the sample from white to cyan above transition temperature is observed from optical microscopic images (Fig. 3(c)). To check whether there is any role of substrate in the observed colour change, we captured optical images of free standing VO$_2$ crystals taken out from the Si substrate at different temperatures (Figure 4). The free standing crystals also undergo the same behavior as observed for the crystal grown on Si substrate, which confirms that optical changes at transition temperature is solely material property and does not depend on substrate. The colour of the mycrocrystals changed from white to cyan above transition temperature again retains their colour from cyan to white below transition temperature.

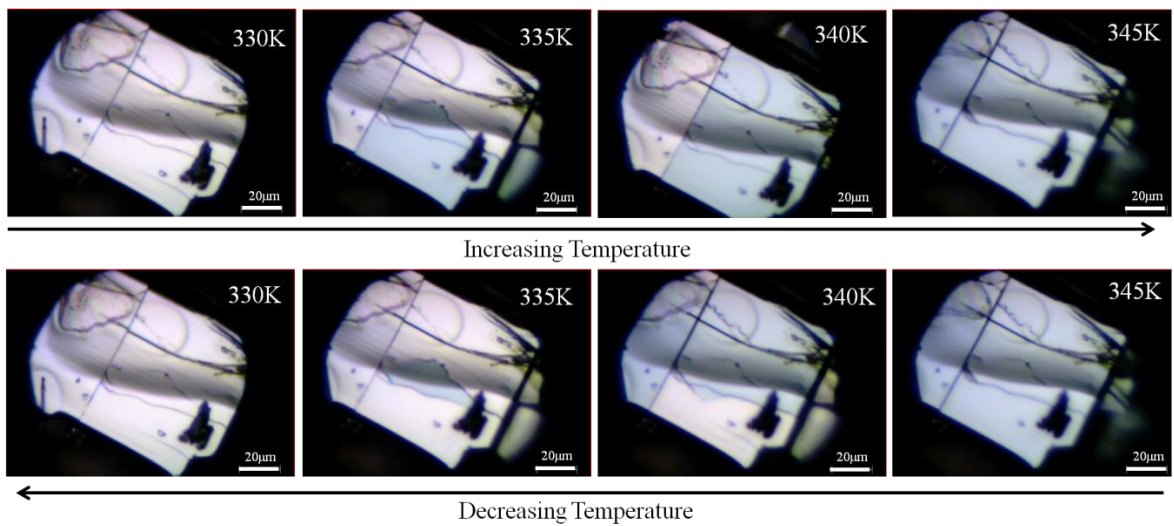

**Figure 4.** Optical images of free standing VO$_2$ micro-crystals with increasing and decreasing temperature.

UV-Vis spectroscopy is one of the useful tools for the determination of colour for transition metal complexes. Transition metal ions can be colored (i.e., absorb visible light) because *d* electrons within the metal atoms can be excited from one electronic state to another [22-23]. For further analysis of the observed change in color during phase transition, temperature dependent UV-Vis absorption spectra were recorded with the help of a light source of excitation wavelength in the range of 200 to 800 nm.

*3.3. Optical properties with the understanding of electronic band structure*

Figure 5(a) shows UV-Vis absorption spectra at different temperature ranging from 300 to 360K. The broad absorption peak ranging the whole visible range for different temperatures of 300 to 340K can be explained as intra-band absorption [24]. A strong absorbance in red region centered at ~ 650 nm is observed above the transition temperature ~350 K. Absorption tail ~700 nm (indicated by solid arrow) is a clear indication of red light absorption above the transition temperature in the metallic phase. The alteration of the upward positive slope in the blue region below 400 nm (indicated by dashed arrow) to negative downward above the transition temperature confirms the reflection of blue light above the phase transition temperature.

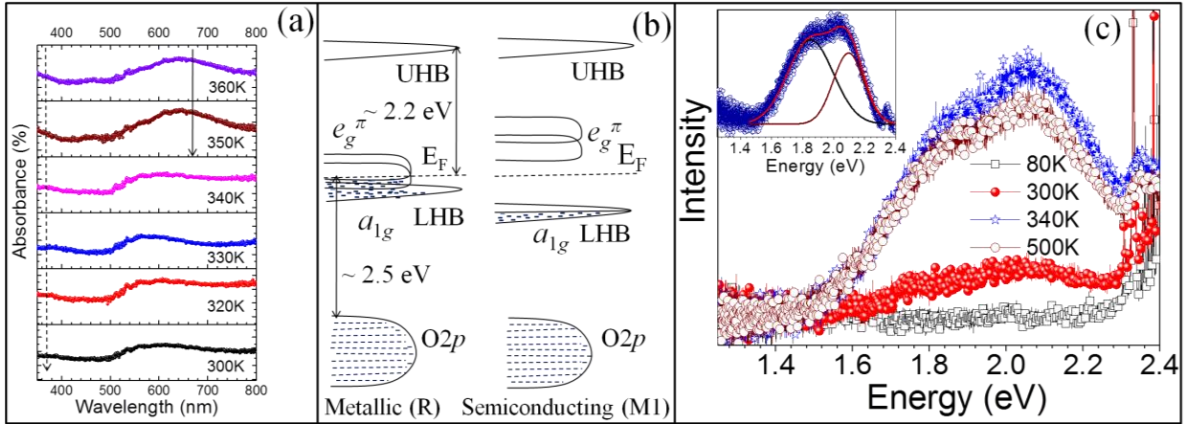

**Figure 5.** (a) UV-Vis absorption spectra of the grown sample with increasing temperature. The dashed line and solid line arrows indicate change in slope and absorption tail, respectively, above the transition temperature. (b) Schematic electronic band structure for semiconducting and metallic phase of $VO_2$ and c) PL spectra of the sample at different temperatures. Inset in (c) shows the deconvoluted PL spectra at 340K.

We try to understand the observed optical properties in the light of electronic transitions for $VO_2$. The electronic band diagrams of $VO_2$ at low temperature insulating and high temperature metallic phases are shown schematically in figure 5(b). As discussed earlier, C-DMFT calculation using electronic correlation (Hubbard $U$) can only explain the opening of an energy gap in the low temperature semiconducting phase of $VO_2$. Thus, it is important to choose a proper value of $U$ to find the correct assessment of the energy gap. The typical value of $U$ for transition metal oxides is ~3 to 4 eV [25]. With the intra-atomic Coulomb repulsion $U$ set to 4.0 eV, Biermann *et al*. found out lower Hubbard band (LHB) about -1.8 eV below the Fermi level and upper Hubbard band (UHB) at about +2.2 eV [17]. The optical absorptions ~2 eV and below are attributed as intraband transition or, free electron absorption in the previous reports [13,14]. We claim that the absorption at red region (~2 eV), as observed from the temperature dependent UV-Vis spectra above the transition temperature, can be ascribed as transition from filled $e_g^\pi$ bands to the UHB. For further analysis, temperature dependent PL spectroscopic studies were performed. Figure 5(c) shows the PL spectra of the as-grown sample in the temperature of 80-500K. At the lowest temperature of 80K no luminescence peak is observed with the laser excitation of 514.5 nm (2.41 eV), which is quite expected, as the valence band of $VO_2$ (O2$p$) is ~2.5 eV below the Fermi level in the semiconducting phase [13,14]. Although, the half-filled lower lying $a_{1g}$ bands (LHB) may fall in the incident energy range, it will not participate in the PL process, as a single electron added to this band will jump to the UHB. The observed low intense broad peak at 300K may be attributed as the emission from UHB to $e_g^\pi$ band as the $e_g^\pi$ band is partially filled at room temperature [26]. At the transition temperature of 340K two broad peaks are observed at 1.85 and 2.09 eV, (inset in Fig, 5(c)) which can be ascribed as transition from filled $a_{1g}$ and $e_g^\pi$ states, respectively, to the UHB which is ~2.2 eV above the Fermi level, $E_F$ [17]. The observation supports the electronic correlation as the driving mechanism for SMT in $VO_2$ [27]. The intensity of the peaks decreases slightly with further increase in temperature up to 500K, which is natural for PL emission with increasing temperature. We have synthesized several samples varying the growth parameters. All of them show PL peaks at same energy value at temperature above the phase transition value only (not shown in figure). However, we failed to measure any PL peak upto the transition temperature starting form 80K, where defect may like to play a role also. This observation confirms that the PL peaks correspond to transition between band states instead of defects. The emission at ~2 eV above the transition temperature supports the absorption of light at red region, as observed from UV-Vis absorption spectroscopic study. As the color of a material depends mostly on three primary colors red, green and blue (R+G+B), the absorption of red light makes the sample cyan (G+B) above the transition temperature as observed from temperature dependent optical microscopic images, captured above the transition temperature (Fig. 3(b)).

## 4. Conclusions

VO$_2$ microcrystals in the monoclinic phase with (011) crystallographic orientation were grown using vapor transport process on Si (111) substrate. The semiconductor to metal phase transition is found out as 340K by disappearance of Raman actives modes in the temperature dependent Raman spectroscopic analysis. A change in color of the VO$_2$ micro crystals from white to cyan is observed with variation of temperature around transition. Temperature dependent UV-Vis spectroscopic analysis shows absorption of red light above the transition temperature, which is further confirmed by emission peaks observed at 1.85 and 2.09 eV in the PL measurements above transition temperature. The absorption of red light is explained from the transition between filled $e_g^\pi$ states and upper Hubbard band (UHB), which supports the electronic correlation as the driving mechanism for SMT in VO$_2$. The observed color change of the VO$_2$ microcrystals from white (R+G+B) to cyan (B+G) is explained by the absorption of red light (R) above the transition temperature and is analyzed with electronic band structure of VO$_2$ for both the semiconducting low and metallic high temperature phases. These thermochromic properties make VO$_2$ applicable as advanced smart windows for overall heat management of a closure.

## Acknowledgments

We thank Arindam Das, P. A. Manoj Kumar and Santanu Parida of SND, IGCAR for valuable suggestions.

**Figure Captions:**

**Figure 1.** (a) FESEM image of as-grown microcrystals and (b) GIXRD spectrum of as-grown sample indicating crystalline (*hkl*) planes of (011) corresponding to M1 phase of $VO_2$

**Figure 2.** Typical Raman spectrum of the as grown $VO_2$ microcrystals with excitation of 514.5 nm at room temperature.

**Figure 3.** (a) Raman spectra of the sample at the temperature range 300 to 360K. Solid and dashed arrows denote increasing and decreasing temperatures, respectively, (b) Reversible temperature dependent resistance measurement for the $VO_2$ crystals showing a change in resistance of four order indicating metal insulator transition with a hysteresis of ~8K (shown by vertical dashed lines). Inset shows linearly fitted (solid line) reversible IV curves at 300K confirming Ohmic behavior (c) optical images of $VO_2$ micro-crystals at different temperatures

**Figure 4.** Optical images of free standing VO2 micro-crystals with increasing and decreasing temperature.

**Figure 5.** (a) UV-Vis absorption spectra of the grown sample with increasing temperature. The dashed line and solid line arrows indicate change in slope and absorption tail, respectively, above the transition temperature. (b) Schematic electronic band structure for semiconducting and metallic phase of $VO_2$ and c) PL spectra of the sample at different temperatures. Inset in (c) shows the deconvoluted PL spectra at 340K.